\documentstyle[prd,preprint,aps]{revtex}
\begin{document}
\tightenlines
\draft
\title{Pairs-Production of Higgs in Association with\\
       Bottom Quarks Pairs at $e^+e^-$ Colliders}

\author{A. Guti\'errez-Rodr\'{\i}guez $^{1, a}$, M. A. Hern\'andez-Ru\'{\i}z $^{2, b}$
        and O. A. Sampayo $^{3}$}

\address{(1) Facultad de F\'{\i}sica, Universidad Aut\'onoma de Zacatecas\\
         Apartado Postal C-580, 98060 Zacatecas, Zacatecas M\'exico.\\
         (a) Cuerpo Acad\'emico de Part\'{\i}culas, Campos y Astrof\'{\i}sica.}

\address{(2) Facultad de Ciencias Qu\'{\i}micas, Universidad Aut\'onoma de Zacatecas\\
         Apartado Postal 585, 98060 Zacatecas, Zacatecas M\'exico.\\
         (b) Cuerpo Acad\'emico de Ciencias B\'asicas de Q.F.B..}

\address{(3) Departamento de F\'{\i}sica, Universidad Nacional del Mar del Plata\\
         Funes 3350, (7600) Mar del Plata, Argentina.}
\date{\today}
\maketitle

\begin{abstract}

In a previous paper, we studied the Higgs pair production in the standard model
with the reaction $e^{+}e^{-}\rightarrow t \bar t HH$. Based on this,
we study the Higgs pair production via $e^{+}e^{-}\rightarrow b \bar b HH$.
We evaluate the total cross section of $b\bar bHH$ and calculate the number
total of events considering the complete set of Feynman diagrams at tree-level,
and compare this process with the process $e^{+}e^{-}\rightarrow t \bar t HH$.
The numerical computation is done for the energy which is expected to be available at a possible Next
Linear $e^{+}e^{-}$ Collider with a center-of-mass energy $800, 1000, 1600$
$GeV$ and luminosity 1000 $fb^{-1}$.
\end{abstract}

\pacs{PACS: 13.85.Lg, 14.80.Bn}


\section{Introduction}

Up to now the Standard Model (SM) \cite{Weinberg} has passed all
accelerator-based experimental tests. It is able to reproduce all experimental
data obtained at high energy $e^+e^-$, $pp$ and $e^\pm p$ colliders. In particular,
the precision data of Large Electron Positron Collider (LEP) has verified the
SM predictions with very high accuracy, and the experimental errors are in
the range of about $0.1 \%-1\%$. On the theoretical side, most of the one-loop
corrections to the prominent observables have been calculated. In some cases,
the leading two-loop corrections are also known. The theoretical errors are
in the $0.1 \%-1\%$ range as well.

While the gauge sector of the SM has been extremely well-tested, our theoretical
ideas about electroweak symmetry breaking are still not completely convincing.
In fact, clarifying the mechanism of electroweak symmetry breaking will be the
central problem we will have to solve with the next generation of high energy
colliders. In the SM, electroweak symmetry breaking is achieved by the Higgs
mechanism. The scalar Higgs boson, also predicted by this mechanism, has not
been found so far.

The search for Higgs bosons is one of the principal missions of present and future
high-energy colliders. The observation of this particle is of major importance
for the present understanding of the interactions of the fundamental particles.
Indeed, in order to accommodate the well established electromagnetic and weak
interaction phenomena, the existence of at least one isodoblete scalar field
to generate fermion and weak gauge bosons masses is required. The SM makes use
of one isodoblete field consisting of three Goldstone bosons among the four
degrees of freedom which are absorbed to build up the longitudinal components
of the massive $W^\pm$, $Z^0$ gauge bosons; one degree of freedom is left over
corresponding to a physical scalar particle, which is the Higgs boson \cite{Higgs}.
Despite its numerous successes in explaining the present data, the SM cannot be
completely tested before this particle has been experimentally observed and its
fundamental properties studied. In particular, the Higgs boson self-interaction. 

In the SM, the profile of the Higgs particle is uniquely determined once its
mass $M_H$ is fixed \cite{Hunter}; the decay width and branching, as well as
the production cross sections, are given by the strength of the Yukawa couplings
to fermions and gauge bosons, which is set by the masses of these particles.
However, the Higgs boson mass is a free parameter and there are two
experimental constraints on this free parameter.

The SM Higgs boson has been searched by LEP in the Higgs-strahlung process
$e^+e^-\to HZ$ for c.m. energies up to $\sqrt{s}=209$ $GeV$ and with a large
collected luminosity. In the summer of 2002, the final results of the four LEP
collaborations were published and some changes were made with respect to the
original publication. In particular, the inclusion of more statistics, the
revision of backgrounds, and the reassessment of systematic errors. When these results are
combined, an upper limit $M_H \geq 114.4$ $GeV$ is established at the $95 \%$
confidence level \cite{Working Group}. However, this upper limit, in the
absence of additional events with respect to SM predictions, was expected
to be $M_H > 115.3$ $GeV$; the reason is that there is a $1.7\sigma$ excess
[ compared to the value $2.9\sigma$ reported at the end of 2000 ] of events
for a Higgs boson mass in the vicinity of $M_H= 116$ $GeV$ \cite{Working Group}.

The second constraint comes from the accuracy of the electroweak observables
measured at LEP, SLAC Large Detector (SLC), and the Fermilab Tevatron,
which provide sensitivity to $M_H$. The Higgs boson contribute logarithmically,
$\propto\log(\frac{M_H}{M_W})$, to the radiative corrections to the $W/Z$
boson propagators. The status, as found in Summer 2002, is summarized in 
Reference \cite{LEPSLD}. When all available data (i.e. the $Z^0$-boson pole LEP and SLC data,
the measurement of the $W$ boson mass and total width, the top-quark mass and
the controversial NuTeV result) is taken into account, one obtains a Higgs boson
mass of $M_H=81^{+42}_{-33}$ $GeV$, leading to a 95$\%$ confidence level upper
limit of $M_H < 193$ $GeV$ \cite {LEPSLD}.

The trilinear Higgs self-coupling can be measured directly in
pair-production of Higgs particles at hadron and high-energy $e^+e^-$ linear
colliders. Higgs pairs can be produced through double Higgs-strahlung of $W$
or $Z$ bosons \cite{Gounaris,Ilyin,Djouadi,Kamoshita}, $WW$ or $ZZ$ fusion
\cite{Ilyin,Boudjema,Barger,Dobrovolskaya,Dicus}; moreover, through
gluon-gluon fusion in $pp$ collisions \cite{Glover,Plehn,Dawson} and
high-energy $\gamma\gamma$ fusion \cite{Ilyin,Boudjema,Jikia} at photon
colliders. The two main processes at $e^+e^-$ colliders are double
Higgs-strahlung and $WW$ fusion:

\begin{eqnarray}
\mbox{double Higgs-strahlung}&:& e^+e^- \to ZHH  \nonumber \\
\mbox{$WW$ double-Higgs fusion}&:& e^+e^- \to \bar\nu_e \nu_e HH.
\end{eqnarray}

The $ZZ$ fusion process of Higgs pairs is suppressed by an order of magnitude
because the electron-Z coupling is small. However, the process $e^+e^-
\rightarrow ZHH$ has been studied \cite{Gounaris,Ilyin,Djouadi,Kamoshita} extensively.
This three-body process is important because it is sensitive to Yukawa couplings.
The inclusion of four-body processes with heavy fermions $f$, $e^{+}e^{-}\rightarrow f\bar f HH$,
in which the SM Higgs boson is radiated by a $b(\bar b)$ quark at future
$e^{+}e^{-}$ colliders  \cite{NLC,NLC1,JLC,A.Gutierrez} with a c.m. energy
in the range of 800 to 1600 $GeV$, as in the case of DESY TeV Energy Superconducting
Linear Accelerator (TESLA) machine \cite{TESLA}, is necessary in order to know its impact on the three-body mode processes and also
to search for new relations that could have a clear signature of the Higgs
boson production.

The Higgs coupling with top quarks, which is the largest coupling in the SM,
is directly accessible in the process where the Higgs boson is radiated
off top quarks, $e^{+}e^{-}\rightarrow t\bar t HH$, followed by the process
$e^{+}e^{-}\rightarrow b\bar b HH$. These processes depend on the
Higgs boson triple self-coupling, which could lead us to obtain the first
non-trivial information on the Higgs potential. We are interested in finding
regions that allow the observation of the process $b\bar bHH$ at the
next generation of high energy $e^{+}e^{-}$ linear colliders. We consider the
complete set of Feynman diagrams at tree-level (Fig. 1) and used the CALCHEP
\cite{Pukhov} packages for the evaluation of the amplitudes and of the cross
section.

This paper is organized as follows: In Sec. II, we present the total cross section
for the process $e^{+}e^{-}\rightarrow b \bar b HH$ at next generation linear
$e^{+}e^{-}$ colliders, and compare it with $e^{+}e^{-}\rightarrow t \bar t HH$.
In Sec. III, we give our conclusions.

\section{Double Higgs Production Cross Section in the SM at Next Generation
         Linear Positron-Electron Colliders}

In this section we present numerical result for $e^{+}e^{-}\rightarrow b \bar b HH$
with double Higgs production and compare it with the process $e^{+}e^{-}\rightarrow t \bar t HH$.
We carry out the calculations using the framework of the Standard Model
at next generation linear $e^{+}e^{-}$ colliders. We used CALCHEP \cite{Pukhov}
packages for calculations of the matrix elements and cross sections. These
packages provide automatic computation of the cross sections and distributions
in the SM as well as their extensions at the tree level. Both processes $e^{+}e^{-}\rightarrow b \bar b HH$
and $e^{+}e^{-}\rightarrow t \bar t HH$ are estimated, including a complete
set of Feynman diagrams for $e^{+}e^{-}\rightarrow b \bar b HH$. We consider
the high energy stage of a possible Next Linear $e^{+}e^{-}$ Collider with
$\sqrt{s}=800, 1000, 1600$ $GeV$ and design luminosity 1000 $fb^{-1}$.

For the SM parameters, we have adopted the following: the Weinber
angle $\sin^2\theta_W=0.232$, the mass ($m_b=4.5$ $GeV$) of the
bottom quark, the mass ($m_t=175$ $GeV$) of the top quark, and 
the mass ($m_{Z^0}=91.2$ $GeV$) of the $Z^0$, having taken the mass
$M_H$ of the Higgs boson as input \cite{Lett}.

In order to illustrate our results of the production of Higgs pairs in the SM,
we present a plot for the total cross section as a function of Higgs
boson mass $M_{H}$ for both processes $e^{+}e^{-}\rightarrow b \bar b HH (t \bar t HH)$
in Fig. 2. We observe in this figure that the total cross section for the double
Higgs production of $b\bar b HH$ and $t \bar t HH$ is of the order of $0.03$ $fb$
for Higgs masses in the lower part of the intermediate range. The cross
sections are at the level of a fraction of femtobarn, and they quickly drop as
they approach the kinematic limit. Under these conditions, it would be very difficult
to extract any useful information about the Higgs self-coupling from the studied
process unless the $e^+e^-$ machine works with very high luminosity.

The cross section for double Higgs boson production in the intermediate mass
range is presented in Figs. 3 and 4 for total $e^+e^-$ energies of
$\sqrt{s}=1000, 1600$ $GeV$. The cross sections are shown for unpolarized
electrons and positrons beams. As in the case shown in Fig. 2, the cross section
is at the level of a fraction of femtobarn and decreases with rising energy
beyond the threshold region. However, the cross section increases with rising
self-coupling in the vicinity of the SM value. The sensitivity to the $HHH$
self-coupling is demonstrated in Ref. \cite{A.Gutierrez1} for $\sqrt{s}= 800$
$GeV$ and $M_H= 130$ $GeV$ by varying the trilinear coupling
$\kappa \lambda_{HHH}$ within the range $\kappa=-1$ and $+2$.

We observe in Figs. 2-4 that the cross section decreases as energy increases,
but still far enough from the threshold. At some given energy, the total
cross sections has its maximum value of $\sigma^{Tot}_{max}$ depending on the
Higgs mass. Since fermion chirality is conserved at the $Z^0-fermion$ vertex,
the cross section may increase by almost double when electrons and positrons
are polarized.

Fig. 5 shows the total cross section as a function of the center-of-mass
energy $\sqrt{s}$ for one representative value of the Higgs mass $M_H=130$
$GeV$. We observe that the cross section is very sensitive to the Higgs
boson mass and decreases when $M_H$ increases. Our conclusion is that
for an intermediate Higgs boson, a visible number of events would be produced,
as illustrated in Table I.

For center-of-mass energies of 800-1600 $GeV$ and high luminosity, the
possibility of observing the processes $b\bar bHH$ and $t\bar tHH$
are promising as shown in Table I. Thus, a high-luminosity $e^+e^-$ linear
collider is a very high precision machine in the context of Higgs physics.
This precision would allow the determination of the complete profile of the
SM Higgs boson, in particular if its mass is smaller than $\sim 130$ $GeV$.

\vspace*{5mm}

\begin{center}
\begin{tabular}{|c|c|c|c|}
\hline
Total Production of Higgs Pairs & \multicolumn{3}{c|}{$e^{+}e^{-}\rightarrow b \bar b HH \hspace{2mm}(t \bar t HH)$}\\
\hline
\cline{2-4} & $\sqrt{s}= $ & $\sqrt{s}= $ & $\sqrt{s}= $  \\
$M_H(GeV)$ & 800 $GeV$ & 1000 $GeV$ & 1600 $GeV$ \\
\hline \hline
 100 & 25 (19) & 18 (28) & 10 (21) \\
 110 & 26 (13) & 19 (22) & 10 (18) \\
 120 & 23 (9)  & 18 (17) & 10 (16) \\
 130 & 23 (6)  & 18 (13) & 10 (14) \\
 140 & 21 (3)  & 18 (19) & 10 (12) \\
 150 & 19 (2)  & 16 (8)  &  9 (11) \\
 160 & 19 (1)  & 16 (6)  &  9 (10) \\
 170 & 17 (-)  & 15 (4)  &  9 (9) \\
 180 & 15 (-)  & 15 (3)  &  8 (8) \\
 190 & 14 (-)  & 14 (2)  &  8 (7) \\
 200 & 12 (-)  & 13 (1)  &  8 (6) \\
\hline
\end{tabular}
\end{center}

\begin{center}
Table I. Total production of Higgs pairs in the SM for ${\cal L}=1000$ $fb^{-1}$
$m_b=4.5$ $GeV$ and $m_t=175$ $GeV$.
\end{center}

In Fig. 6, we also include a contours plot for the number of events of the
studied processes, as a function of $M_H$ and $\sqrt{s}$. These contours are
obtained from Table I.

Although the Higgs coupling with top quarks, the largest coupling in
the SM, is directly accessible in the process where the Higgs boson is radiated
off top quarks $e^{+}e^{-}\rightarrow t \bar t HH$. The coupling with bottom
quarks is also accessible in the reaction where the Higgs is radiated by a
$b(\bar b)$ quark, $e^{+}e^{-}\rightarrow b \bar b HH$. For $M_H\lesssim 130$ $GeV$,
the Yukawa coupling can be measured with a precision of less than $5 \%$ at
$\sqrt{s}= 800$ $GeV$ with a luminosity of ${\cal L}=1000$ $bf^{-}$.

Finally, the measurement of the trilinear Higgs self-coupling, which is the
first non-trivial test of the Higgs potential, is accessible in the double
Higgs production processes $e^{+}e^{-}\rightarrow b \bar b HH$ and in the
$e^{+}e^{-}\rightarrow t \bar t HH$ process at high energies. Despite its
smallness, the cross sections can be determined with an accuracy of the order
of $20 \%$ at a 800 $GeV$ collider if a high luminosity (${\cal L}=1000$ $fb^{-}$)
is available.

\section{Conclusions}

In conclusion, the double Higgs production, in association with
$b(\bar b)$ and $t(\bar t)$ quarks ($e^{+}e^{-}\rightarrow b \bar b HH, t \bar t HH$),
will be observable at the Next Generation Linear $e^+e^-$ Colliders. The
study of these processes is important in order to know their impact on the
three-body process and could be useful to probe anomalous $HHH$
coupling given the following conditions: very high luminosity,
excellent $b$ tagging performances, center-of-mass large energy, and
intermediate range Higgs boson mass.

\vspace{2.5cm}


\begin{center}
{\bf Acknowledgments}
\end{center}

This work was supported in part by SEP-CONACyT (Proyects: 2003-01-32-001-057, 40729-F),
Sistema Nacional de Investigadores (SNI) (M\'exico). O.A. Sampayo would like
to thank CONICET (Argentina). We also would like to thank Maureen Harkins
(cetet@uaz.edu.mx) for proofreading of the text.

\newpage

\begin{center}
{\bf FIGURE CAPTIONS}
\end{center}

\vspace{5mm}

\bigskip

\noindent {\bf Fig. 1} Feynman diagrams at tree-level for $e^{+}e^{-}
\rightarrow b\bar b HH$.

\bigskip

\noindent {\bf Fig. 2} Total cross section of the Higgs pairs production
          $e^{+}e^{-}\rightarrow b\bar b HH (t\bar t HH)$ as function of the Higgs
          boson mass $M_H$ for $\sqrt{s} =800$ $GeV$ with $m_{b} = 4.5$ $GeV$ and
          $m_{t} = 175$ $GeV$.

\bigskip

\noindent {\bf Fig. 3} The same as in Fig. 2, but for $\sqrt{s} =1000$ $GeV$.

\bigskip

\noindent {\bf Fig. 4} The same as in Fig. 2, but for $\sqrt{s} =1600$ $GeV$.

\bigskip

\noindent {\bf Fig. 5} Total cross section of the Higgs pairs
          production $e^{+}e^{-}\rightarrow b\bar b HH (t\bar t HH)$ as a function
          of the center-of-mass energy $\sqrt{s}$ for one representative value
          of the Higgs mass $M_H=130$ $GeV$ with $m_b=4.5$ $GeV$ and $m_t=175$
          $GeV$.

\bigskip

\noindent {\bf Fig. 6} Contours plot for the number of events of both processes
          $e^{+}e^{-}\rightarrow b\bar b HH $ and $e^{+}e^{-}\rightarrow t\bar t HH$
          as a function of $M_H$ and $\sqrt{s}$.


\newpage

\begin{references}

\bibitem{Weinberg} S. Weinberg, Phys. Rev. Lett. {\bf 19}, 1264 (1967);
                   A. Salam, in {\it Elementary Particle Theory}, ed. N. Southolm (Almquist and
                   Wiksell, Stockholm, 1968), p.367; S.L. Glashow, Nucl. Phys. {\bf 22}, 257 (1967).

\bibitem{Higgs} P. W. Higgs, Phys. Rev. Lett. {\bf 12}, 132 (1964);
                Phys. Rev. Lett. {\bf 13}, 508 (1964);
                Phys. Rev. Lett. {\bf 145}, 1156 (1966);
                F. Englert, R. Brout, Phys. Rev. Lett. {\bf 13}, 321 (1964);
                G. S. Guralnik, C. S. Hagen, T. W. B. Kibble, Phys. Rev. Lett. {\bf 13},
                585 (1964).

\bibitem{Hunter} For a review on the Higgs sector in the SM, see: J. F. Gunion,
                 H. E. Haber, G. L. Kane and S. Dawson; "The higgs Hunter's Guide",
                 Addison-Wesley, Reading  1990.


\bibitem{Working Group} The LEP Higgs Working Group, Note/2002-01 for the SM;
                        The LEP Higgs Working Group, hep-ex/0107029; hep-ex/0107030.

\bibitem{LEPSLD}  The LEP and SLD Electroweak Working Group, LEPEWW/2002-02
                  (Dec. 2002), hep-ex/0212036; hep-ex/0112021.

\bibitem{Gounaris} G. Gounaris, D. Schildknecht and F. Renard, Phys. Lett.
                   {\bf 83B}, 191 (1979); {\bf 89B}, 437(E) (1980);
                   V. Barger, T. Han and R.J.N. Phillips, Phys. Rev. {\bf D38}, 2766 (1988).

\bibitem{Ilyin} V. A. Ilyin, A. E. Pukhov, Y. Kurihara, Y. Shimizu and T.
                Kaneko, Phys. Rev. {\bf D54}, 6717 (1996).

\bibitem{Djouadi} A. Djouadi, H. E. Haber and P. M. Zerwas, Phys. Lett.
                  {\bf B375}, 203 (1996); A. Djouadi, W. Kilian, M. M. Muhlleitner and P. M.
                  Zerwas, Eur. Phys. J. {\bf C10}, 27 (1999); P. Oslan, P. N. Pandita, Phys.
                  Rev. {\bf D59}, 055013 (1999); F. Boudjema and A. Semenov, hep-ph/0201219;
                  A. Djouadi, hep-ph/0205248; Abdelhak Djouadi,
                  hep-ph/0503172, and references therein.

\bibitem{Kamoshita} J. Kamoshita, Y. Okada. M. Tanaka and I. Watanabe,
                    hep-ph/9602224; D. J. Miller and S. Moretti, hep-ph/0001194; D. J. Miller
                    and S. Moretti, Eur. Phys. J. {\bf C13}, 459 (2000).

\bibitem{Boudjema} F. Boudjema and E. Chopin, Z. Phys. {\bf C73}, 85 (1996).

\bibitem{Barger} V. Barger and T. Han, Mod. Phys. Lett. {\bf A5}, 667 (1990).

\bibitem{Dobrovolskaya} A. Dobrovolskaya and V. Novikov, Z. Phys. {\bf C52}, 427 (1991).

\bibitem{Dicus} D. A. Dicus, K. J. Kallianpur and S. S. D. Willenbrock,
                Phys. Lett. {B200}, 187 (1988);  A. Abbasabadi, W. W. Repko, D. A. Dicus and
                R. Vega, Phys. Rev. {\bf D38}, 2770 (1988);  Phys. Lett. {\bf B213}, 386 (1988).

\bibitem{Glover} E. W. N. Glover and J. J. van der Bij, Nucl. Phys. {\bf B309}, 282 (1988).

\bibitem{Plehn} T. Plehn, M. Spira and P. M. Zerwas, Nucl. Phys. {\bf B479},
                46 (1996); Nucl. Phys. {\bf B531}, 655 (1998).

\bibitem{Dawson} S. Dawson, S. Dittmaier and M. Spira, Phys. Rev. {\bf D58},
                 115012 (1998).

\bibitem{Jikia}  G. Jikia, Nucl. Phys. {\bf B412}, 57 (1994).

\bibitem{NLC} NLC ZDR Desing Group and the NLC Physics Working Group, S.
              Kuhlman {\it et al.}, {``Physics and Technology of the Next Linear Collider"},
              hep-ex/9605011.

\bibitem{NLC1} The NLC Design Group, C. Adolphsen {\it et al.} {``Zeroth-Order
               Design Report  for the Next Linear Collider"}, LBNL-PUB-5424, SLAC
               Report No. 474,  UCRL-ID-124161 (1996).

\bibitem{JLC} JLC Group, JLC-I, KEK Report No. 92-16, Tsukuba (1992).

\bibitem{A.Gutierrez} A. Guti\'errez-Rodr\'{\i}guez, M. A. Hern\'andez-Ru\'{\i}z and O. A. Sampayo,
                      Phys. Rev. {\bf D67}, 074018 (2003).

\bibitem{TESLA} TESLA Technical Desing Report, Part III, DESY-01-011C, hep-ph/0106315.

\bibitem{Pukhov} {``COMPHEP - A package for evaluation of Feynman diagrams and
               integration over multi-particle phase space"}, A.Pukhov et al.,
               Preprint INP MSU 98-41/542, hep-ph/9908288.

\bibitem{Lett} Particle Data Group, S. Eidelman {\it et al.}, Phys. Lett. {\bf B592}, 1 (2004).

\bibitem{A.Gutierrez1} A. Guti\'errez-Rodr\'{\i}guez, M. A. Hern\'andez-Ru\'{\i}z and O. A. Sampayo,
                       in preparation.

\end{references}
\end{document}